\documentclass[10pt]{bmc_article}

\usepackage{cite} 
\usepackage{url}  
\usepackage{ifthen}  
\usepackage{multicol}   
\usepackage[utf8]{inputenc} 
\def\includegraphics{}

\newboolean{publ}

\newenvironment{bmcformat}{\begin{raggedright}\baselineskip20pt\sloppy\setboolean{publ}{false}}{\end{raggedright}\baselineskip20pt\sloppy}

\begin{document}
\begin{bmcformat}


\title{Identification and quantification of Granger causality between gene sets}
 

\author{Andr\'e Fujita\correspondingauthor$^{1}$%
       \email{Andr\'e Fujita\correspondingauthor - andrefujita@riken.jp}%
      \and
         Jo\~ao Ricardo Sato$^2$%
         \email{Jo\~ao Ricardo Sato - joao.sato@ufabc.edu.br}
       \and
       Kaname Kojima$^3$%
       \email{Kaname Kojima - kaname@ims.u-tokyo.ac.jp}
       \and
       Luciana Rodrigues Gomes$^4$%
       \email{Luciana Rodrigues Gomes - lugomes@iq.usp.br}
			 \and
       Masao Nagasaki$^3$%
       \email{Masao Nagasaki - masao@ims.u-tokyo.ac.jp}
       \and
       Mari Cleide Sogayar$^4$%
       \email{Mari Cleide Sogayar - mcsoga@iq.usp.br}
       and 
         Satoru Miyano$^{1,3}$%
         \email{Satoru Miyano - miyano@ims.u-tokyo.ac.jp}%
      }


\address{%
    \iid(1)Computational Science Research Program, RIKEN, 2-1 Hirosawa, Wako, Saitama, 351-0198, Japan.\\
    \iid(2)Center of Mathematics, Computation and Cognition, Universidade Federal do ABC, Rua Santa Ad\'elia, 166 - Santo Andr\'e, 09210-170, Brazil.\\
    \iid(3)Human Genome Center, Institute of Medical Science, University of Tokyo, 4-6-1 Shirokanedai, Minato-ku, Tokyo, 108-8639, Japan.\\
    \iid(4)Chemistry Institute, University of S\~ao Paulo, Av. Lineu Prestes, 748 - S\~ao Paulo, 05508-900, Brazil.
}%

\maketitle


\begin{abstract}
        \paragraph*{Background:} Wiener and Granger have introduced an intuitive concept of causality (Granger causality) between two variables which is based on the idea that an effect never occurs before its cause. Later, Geweke has generalized this concept to a multivariate Granger causality, i.e., $n$ variables Granger-cause another variable. Although Granger causality is not ``effective causality'' in the Aristothelic sense, this concept is useful to infer directionality and information flow in observational data. Granger causality is usually identified by using VAR (Vector Autoregressive) models due to their simplicity. In the last few years, several VAR-based models were presented in order to model gene regulatory networks. Here, we generalize the multivariate Granger causality concept in order to identify Granger causalities between {\it sets} of gene expressions, i.e., whether a set of $n$ genes Granger-causes another set of $m$ genes, aiming at identifying and quantifying the flow of information between gene networks (or pathways). 
      
        \paragraph*{Results:} The concept of Granger causality for {\it sets} of variables is presented. Moreover, a method for its identification with a bootstrap test is proposed. This method is applied in simulated and also in actual biological gene expression data in order to model regulatory networks.

        \paragraph*{Conclusions:} This concept may be useful to understand the complete information flow from one network or pathway to the other, mainly in regulatory networks. Linking this concept to graph theory, sink and source can be generalized to node sets. Moreover, hub and centrality for sets of genes can be defined based on total information flow. Another application is in annotation, when the functionality of a set of genes is unknown, but this set is Granger caused by another set of genes which is well studied. Therefore, this information may be useful to infer or construct some hypothesis about the unknown set of genes.
\end{abstract}

\ifthenelse{\boolean{publ}}{\begin{multicols}{2}}{}


\section*{Background}
	Norbert Wiener \cite{Wiener56} conceived the notion that, if the prediction of one time series could be statistically improved by incorporating the knowledge of past values of a second one, then the latter is said to have a ``causal'' influence on the former. Clive Granger \cite{Granger69} has formalized the prediction idea in the context of VAR (Vector Autoregressive) models. More specifically, if the variance of the autoregressive prediction error of one time series at the present time is statistically reduced by inclusion of past measurements from another time series, then the latter is said to have a Granger causal influence on the former. From this definition it is clear that the flow of time plays a crucial role in allowing inferences to be made about directional causal influences from time series data. Due to its simplicity and intuitive idea that an effect never occurs before its cause, it has been widely used in several areas such as econometrics \cite{McCrorie06, Wong06, McCracken07}, neuroscience \cite{Brovelli04, Roebroeck06, Sato06} and more recently, in bioinformatics \cite{Mukhopadhyay07, Fujita07a, Fujita07b, Fujita08, Kojima08, Nagarajan08, Nagarajan09}. The idea is that, if a variable $x$ affects a variable $y$, past values of the former should be useful in generating predictions for the latter variable.\pb

	Later, Geweke \cite{Geweke84} has generalized the bivariate Granger causality to a multivariate fashion in order to identify conditional Granger causality. To illustrate the bivariate and multivariate Granger causalities, suppose three processes, where one drives the other two with differential time delays. A pairwise analysis would indicate a causal influence from the process that receives an early input to the process that receives a late input. Conditional Granger causality may be useful to disambiguate these situations, since it has the ability to resolve whether the interaction between two time series is direct or is mediated by another recorded time series and whether the causal influence is simply due to differential time delays in their respective driving inputs.\pb

	Nagarajan and Upreti (2008) \cite{Nagarajan08} and Nagarajan (2009) \cite{Nagarajan09} investigated the use of bivariate VAR for acyclic approximations of networks composed of two genes by exploring the parameters defined as transcriptional noise variance, autoregulatory feedback, and transcriptional coupling strength, which may influence some measures of Granger Causality. These authors have shown that under some conditions, uni-directional/acyclic approximations may provide meaningful and useful results.\pb

	Frequently, conditional Granger causality is identified by using VAR models due to its simplicity \cite{Lutkepohl06}. In the last few years, several extensions of the standard VAR model namely, Dynamic VAR (DVAR) \cite{Fujita07a}, Sparse VAR (SVAR) \cite{Fujita07b, Opgen-Rhein07, Shimamura09, Lozano09} and Nonlinear VAR (NVAR) \cite{Fujita08, Chen04, Guo08, Marinazzo08a, Marinazzo08b} have been proposed to model gene regulatory networks. DVAR were developed in order to identify time-varying Granger causalities, i.e., to identify structural changes along time (cell cycle) in regulatory networks. However, DVAR identifies only linear Granger causalities. In order to overcome this limitation, a model capable to identify nonlinear Granger causalities, namely, NVAR, was introduced. Another problem in bioinformatics is the high dimensional characteristic of gene expression data, where the number of parameters (genes) is higher than the number of observations (microarrays). Therefore, constructing regulatory networks with statistical tests are difficult. In order to identify linear Granger causalities in this context, SVAR was proposed \cite{Fujita07b, Opgen-Rhein07, Shimamura09}.\pb
	
	These models are useful to identify Granger causalities between genes (gene expression signals), however, sometimes, it is necessary to identify and quantify Granger causality between {\it sets} of genes. Thus, for example, one may need to quantify the sum of Granger causalities from one gene cluster to the other gene cluster in order to study the total amount of information flow between different regulatory pathways and to identify which pathway is crucial. Another application may be the identification of Granger causality between a set of genes whose biological function is yet unknown and another set of genes which are better studied and for which more characteristics are known. This latter case may help to determine and infer the possible biological process affected by the set of genes whose functionalities are not yet known.\pb
	
	The theoretical generalization of Granger causality between {\it sets} of variables has not been sufficiently explored. Here, we propose a definition of Granger causality between {\it sets} of time series, where $m$ time series Granger-cause $n$ other time series, thus generalizing the two previous definitions (bivariate and multivariate). Furthermore, a method for identification of Granger causality and a statistical test based on nonparametric bootstrap are also proposed. The results are illustrated by simulations and, also, by application of this new concept to actual biological gene expression data.

\section*{Results and Discussion}
  \subsection*{Simulation}
	In order to study the properties of our method to identify Granger causalities between sets of genes, we have designed a simulation study which contains 14 artificial genes in a time series of length $T=100$ (simulation 1).\pb
	
	Figure 1A illustrates an example of time series obtained by our simulation study. Figure 1B illustrates the structure of the artificial regulatory network and the interaction between sets (I, II and III). Figure 1C represents the true model, i.e., if our approach is working correctly, it must identify the structure illustrated in Figure 1C.\pb
	
	Notice in Table 1 that where there is no edge, the number of false-positives is actually controlled to $5\%$ ($\sim 500$). In bold, the edges which actually are present in the network are illustrated. Notice that, as expected, where there is an edge, the number of identifications is higher than where there is no edge. Moreover, verify that the number of false-positives is actually controlled to 5\%, i.e., the bootstrap procedure is working adequately.\pb

	Interpreting the loops (auto-regressions) from I to I and from II to II, they are representing the total information flow contained in networks I and II, respectively. Another interpretation for the loops may be the density of the network, since this density may represent the {\it sum} of the information flow. On the other hand, no Granger causality is found in set III, therefore, no loop is present. It is necessary to point out that these loops are exactly the $CCA({\bf Y}_{t}^{i},{\bf Y}_{t-1}^{i}|{\bf Y}_{t}\backslash \{{\bf Y}_{t-1}^{i}\})$ (CCA: Canonical Correlation Analysis), i.e., the total connectivity inside set $Y^{i}$, where $i$=I, II or III. Notice that CCA is always applied between the past of one subset and the present of another subset of the time series.\pb

	Figure 2A illustrates the structure of an artificial network and the interaction between sets of time series (Figure 2B) described in simulation 2. In this simulation 2, both approaches, based on CCA and blockwise VAR proposed by \cite{Wang06} were applied in order to evaluate the power of the proposed method. The blockwise VAR model was applied using all the variables, i.e., a large network was constructed using VAR. The test between sets for VAR were performed by testing the estimated coefficients between sets using the Wald's test (whether all of them are equal to zero simultaneously). The Wald's test is equivalent to test the ratio of the prediction's error variance proposed by \cite{Geweke84}. Notice, in Table 2 that, again, the false-positives are actually controlled to 5\%. In bold, are the number of times Granger causality was identified in a total of 10,000 simulations. Analyzing the number of Granger causalitites identified by both methods, the method based on CCA is more powerful than the standard VAR. This is due to the CCA characteristic which maximizes the correlation by calculating the optimum linear combination of both sets while the blockwise VAR sets equal weights to each time series. In a biological sense, time series $W_{3}$ (Figure 2) may be interpreted as a transcription factor which Granger causes several other genes. The method based on CCA is also clearly more powerful in the case of auto loops (from I to I, from II to II and from III to III).

	\subsection*{Biological data}
	In order to illustrate a practical application of the proposed model, an actual biological regulatory network was modeled. The following 15 genes, namely, RECK, SRC, C-MYC, TIMP2, TP53, p21, GADD45A, FGFRL1, FGF2, FGFR2, FGFR1, NEMO, IKKA, IKB and NFKB, were selected, the same which were used in our previous works \cite{Fujita07a, Fujita07b, Fujita08}. The proposed approach was applied to the analysis of HeLa cell cycle gene expression data collected by \cite{Whitfield02}. HeLa is an immortal cell line, derived from a human epithelial cervical carcinoma. Its cell cycle is of approximately 16 hours. The gene expression data used in this study contains three complete cell cycles, i.e., 48 time points distributed at intervals of one hour. The HeLa gene expression data is freely available at: \cite{Hela}.\pb
	
	The sets of genes were composed as follows: (I) RECK, SRC, C-MYC and TIMP2; (II) TP53, p21 and GADD45A; (III) FGFRL1, FGF2, FGFR2 and FGFR1 and (IV) NEMO, IKKA, IKB and NFKB. The networks illustrated inside each cluster (Figure 3) were obtained by applying the standard VAR model. The edge links between the clusters were obtained by the method proposed here. Unfortunately, the auto-loops in sets III and IV were not identified probably due to the power of the test. Increasing time series length may increase the power of the test and, consequently, identifying the auto-loops. Notice that in the simulations, the proposed approach was able to correctly identify the Granger causalities.\pb
	
\indent The tumor progression processes, as other pathological and physiological events, are regulated by responses to changes in the external environment, and by following coordinated modulation of gene expression in order to appropriately adapt to these changes. Several molecules are responsible for controlling cell behavior, among which, growth and transcriptional factors display prominent functions in the context of tumorigenesis. The networks of FGF, c-myc, p53 and NFKB, used in this work to demonstrate a biological application of the proposed model, are able to control almost all processes involved in development and tumor progression. Proliferation and differentiation, cell survival and death, cell division and cycle arrest, cell migration, adhesion and invasion are just some examples of cellular phenotypes modulated by these molecules \cite{Huang02}. Therefore, the choice of these sets of genes is appropriate for validating the biological application of the proposed model. Thus, prediction of the interaction between these important gene clusters could elucidate the complex relationship among them and, consequently, allow construction of hypotheses about how the interaction between these pathways orchestrate the regulation of processes that are often mutually exclusive.\pb

	Several members of the fibroblast growth factors (FGFs) family and their respective specific cell surface receptors (FGFRs) play important roles in a variety of normal and pathological processes, including tumor transformation. In the tumor context, FGFs are well known for promoting tumor angiogenesis and invasiveness, and for acting as mitogenic agents and cell death inhibitors in many cell types \cite{Kwabi-Addo08}. However, recent reports have shown that some of these molecules can also promote tumor suppression, depending on the cell type and the number and strength of stimuli. FGF2, a gene belonging to set III, is also implied in restoring tumor defense mechanisms in malignant cells \cite{Costa08}.\pb

	By controlling a variety of downstream target genes, the p53 transcription factor has many tumor-suppressor activities. The p53 protein is modified in more than 50\% of human tumors. Thus, many processes are subverted due to p53-alteration, such as, for instance: cell cycle arrest, inhibition of metastasis and angiogenesis, DNA repair and cell death induction \cite{Jin01}. On the other hand, aberrant c-myc activity is associated with genomic instability and tumor progression. This transcriptional factor is encoded by a proto-oncogene, whose product modulates transcriptional processes during normal cell growth and proliferation \cite{Wade06}. However, in the tumoral context, deregulated c-myc activity is a classic molecular marker of poor prognostics in several types of cancers. Moreover, the c-Myc protein is able to regulate other important events involved in tumor transformation, including hormone dependence, invasiveness and metastatic potential \cite{Shiu93}. In all of these processes, c-Myc positively associates with tumor progression. Like c-Myc, the NFKB family of transcription factors has been shown to be constitutively activated in various human malignancies, namely: leukemias, lymphomas, and a number of solid tumors. As the other three pathways analyzed in this work, NFKB is able to control a variety of events by regulating the expression of genes involved in angiogenesis, metastasis, cell growth, proliferation and apoptosis. Therefore, NFKB has been implicated in transcriptional upregulation of several growth factors, cytokines, adhesion molecules, antiapoptotic proteins and oncogene products \cite{Chen08}.\pb
	
	Here, we were able to identify the Granger causality between the previously summoned networks. We predicted that the set of genes (I) Granger-causes (II), and that (IV) Granger-causes (II) and (III). Therefore, by applying our model, we predicted that the c-Myc network is able to induct p53 related genes, and that the NFKB pathway positively regulates the FGF and p53 sets of genes (Figure 3). By applying the standard VAR model and jointly testing the coefficients \cite{Geweke84}, it was only possible to identify Granger causality from clusters IV to II, confirming the simulation's results (simulation II) that the power of the proposed method in actual biological data is superior to the traditional one.\pb

	Previous reports have demonstrated that p53-induced G1/S cell-cycle arrest is attributed mainly to its ability to the transcriptionally upregulate p21 and repress c-myc \cite{Brown07}. Induction of p21 by p53 is in accordance with the predicted connectivity previously reported \cite{Fujita07a, Fujita07b, Fujita08}. However, the well established p53, acting as a c-myc inhibitor contradicts the identified Granger causality. On the other hand, there are other described mechanisms demonstrating that c-myc is able to indirectly induce p53 through {\it Arf} regulation. Thus, p14$^{\textrm{ARF}}$ (the {\it Arf} product in humans) inhibits Mdm2 and, consequently, stabilizes p53 \cite{Baudino03}. Therefore, this last mechanism demonstrates that the anticipated relationship of p53 induction by c-myc, predicted by our model, is biologically relevant. Several reports linking NFKB and p53 networks are available. As mentioned for the previous sets of genes, these transcriptional factors can reciprocally regulate each other. NFKB has been reported to induce Mdm2, consequently inhibiting p53 \cite{Tergaonkar02}. In contrast, NFKB has also been described as an inducer of the p53 promoter \cite{Hellin98}. Thus, the positive modulation of the p53 pathway by NFKB set of genes, determined by our model, corroborates previously described mechanisms. Moreover, these results suggest the occurrence of a synergistic mechanism to induce p53, in view of the fact that both c-myc and NFKB networks are involved in up-regulation of this transcriptional factor.\pb

	By examining the sign and the magnitude of the canonical weight assigned to each variable, i.e., the eigenvectors ${\bf a}$ and ${\bf b}$ (see Methods section), it is possible to verify that variables with larger weights contribute more to the variables. Moreover, variables whose weights have opposite signs exhibit an inverse relationship with each other, and variables with weights of the same sign exhibit a direct relationship.

\section*{Conclusions}
  Comparison between this method here proposed and other approaches such as Bayesian networks, is difficult, since, by definition, they analyze/identify different kinds of causality. Moreover, although our previous reports (DVAR, SVAR and NVAR) identify time-varying or nonlinear Granger causalities, they are limited to the usual ``between time series (genes)''. The method we propose here generalizes the standard Granger causality to relationships between {\it sets} of time series, which, to the best of our knowledge, is the first to address this issue.\pb

	Ordinarily, when the number of genes is relatively large, interpreting the pairwise Granger causalities is hopeless. Moreover, linear combinations of variables are often interesting and useful for predictive or comparative purposes. In addition, pairwise Granger causality does not distinguish direct from indirect causalities because it is not conditioned by a third set of time series. The purpose of the definition we present for Granger causality is to summarize the associations between the $Y_{t}^{i}$ and $Y_{t}^{j}$ sets in terms of {\it few} carefully chosen covariances. This summarized measure illustrates the strength of Granger causality between sets $Y_{t}^{i}$ and $Y_{t}^{j}$. This may be useful to understand the complete information flow from one network or pathway to the other, mainly in regulatory networks, where the number of genes is high and interpreting gene-gene Granger causalities may be cumbersome. Linking this concept to graph theory, some graph theoretical properties, such as sink and source can be generalized to node sets. Moreover, hub and centrality for sets of genes can be defined based on its total information flow. Another application is in annotation, i.e., when the functionality of a set of genes is unknown, but this set is Granger caused by another set of genes which is well studied. Therefore, this information may be useful to infer or construct some hypothesis about the unknown set of genes.\pb

	Here, we have assumed that the sets of time series are given. However, in practice, it is necessary to identify which time series belongs to each set. To this purpose, biological {\it a priori} information may be used to define the different gene sets. Thus, for example, if it is known that one set of genes respond to a specific drug and there is another set of genes which respond to the same drug, but no literature information is available about the existence of a pathway between these two sets, one may use our approach in order to obtain some clues about the existence (or not) of this pathway. Another application may be the search for crosstalks between under-studied pathways. On the other hand, an objective and systematic method based on clustering may be used. The latter is the object of our future studies.
 
\section*{Methods}
  \subsection*{Granger causality}
	In order to formalize the concept of Granger causality between {\it sets} of time series, suppose that $\Im_{t}$ is a set containing all relevant information available up to and including time-point $t$. Let $Y_{t}$, $Y_{t}^{i}$ and $Y_{t}^{j}$ be sets of time series containing $k$, $m$ and $n$ time series, respectively, where $Y_{t}^{i}$ and $Y_{t}^{j}$ are disjoint subsets of $Y_{t}$, i.e., each time series only belongs to one set, and thus, $k \geq m+n$. Let  $Y_{t}(h|\Im_{t})$ be the optimal (i.e., the one which produces the minimum mean squared error (MSE) prediction) $h$-step predictor of the set of $m$ time series ${\bf Y}_{t}^{i}$ from the time point $t$, based on the information in $\Im_{t}$. The forecast MSE of the linear combination of $Y_{t}^{i}$ will be denoted by $\Omega_{Y}(h|\Im_{t})$. The set of $n$ time series ${\bf Y}_{t}^{j}$ is said to Granger-cause the set of $m$ time series ${\bf Y}_{t}^{i}$ if

\begin{equation}
	\Omega_{Y}(h|\Im_{t}) < \Omega_{Y}(h|\Im_{t} \backslash \{{\bf Y}_{s}^{j}|s\leq t\}) \indent \textrm{for at least one $h=1,2,\ldots$,}
\end{equation}

	where $\Im_{t} \backslash \{{\bf Y}_{s}^{j}|s\leq t\}$ is the set containing all relevant information except for the information in the past and present of ${\bf Y}_{t}^{j}$. In other words, if ${\bf Y}_{t}^{i}$ can be predicted more accurately when the information in ${\bf Y}_{t}^{j}$ is taken into account, then ${\bf Y}_{t}^{j}$ is said to be Granger-causal for ${\bf Y}_{t}^{i}$.\pb

	Applying the idea of Granger causality to regulatory networks, a set of gene expression time series ${\bf Y}_{t}^{j}$ Granger-causes another set of gene expression time series ${\bf Y}_{t}^{i}$, if linear combinations of ${\bf Y}_{t}^{j}$ provide statistically more significant information about future values of linear combinations of ${\bf Y}_{t}^{i}$ than considering only the past values of ${\bf Y}_{t}^{i}$. Thus, past gene expression values of ${\bf Y}_{t}^{j}$ allow the prediction of more accurate gene expression values of ${\bf Y}_{t}^{i}$. Notice that since this relationship is not reciprocal, Granger causality may be interpreted as information flow \cite{Baccala01}. Moreover, it is important to highlight that Granger causality is not actually inferring ``effective'' causality in the Aristothelic sense, i.e., interaction of gene products (or protein-protein interactions), since the former is based solely on prediction and quantitative criteria, as described before, however, this concept may be useful to suggest some insights on molecular interactions which may then be experimentally tested.\pb

	Notice that this definition generalizes not only the original bivariate Granger causality (where ${\bf Y}_{t}^{i}$ and ${\bf Y}_{t}^{j}$ are 1-dimensional) but, also, the multivariate Granger causality (where ${\bf Y}_{t}^{j}$ is $m$-dimensional and ${\bf Y}_{t}^{i}$ is 1-dimensional), i.e., the bivariate and multivariate cases are special cases of the Granger causality for {\it sets} of time series.

  \subsection*{Identification}
    Due to the simplicity of notation and concepts, only the identification of Granger causality in an Autoregressive (AR) process of order one will be presented, generalization for other orders being straightforward.\pb
    
	Hotelling \cite{Hotelling35, Hotelling36} was the first to tackle the problem of identifying and measuring relationships between two {\it sets} of variables, introducing the Canonical Correlation Analysis (CCA) \cite{Yamanishi03, Revell08, Waaijenborg07, Guo06}. Canonical correlation analysis can be understood as the bivariate correlation of two synthetic variables which are the linear combinations of the two sets of observed variables. The original variables of each set are linearly combined to produce pairs of synthetic variables which have maximal correlation.\pb
	
	The present problem consists in verifying whether ${\bf Y}_{t}^{j}$ Granger-causes ${\bf Y}_{t}^{i}$. In the linear case, Granger causality for {\it sets} of variables may be identified based on the idea initially proposed by Hotelling. Notice that linear combinations provide simple summarized measures of a set of variables.\pb
	
	Therefore, for the linear case, let ${\bf Y}_{t}^{i}$ ($m$-dimensional) and ${\bf Y}_{t}^{j}$ ($n$-dimensional) be two disjoint subsets of ${\bf Y}_{t}$, where ${\bf Y}_{t}$ is a $k$-dimensional set of stationary time series ($k \geq m+n$), in other words, each time series only belongs to one cluster. Then, ${\bf Y}_{t}^{j}$ is Granger non-causal for ${\bf Y}_{t}^{i}$ if the following condition holds:

\begin{equation}
	CCA({\bf Y}_{t}^{i},{\bf Y}_{t-1}^{j}|{\bf Y}_{t}\backslash \{{\bf Y}_{t-1}^{j}\})=\rho=0
\end{equation}

where $\rho$ is the largest correlation calculated by Canonical Correlation Analysis. Notice that CCA is applied to the time lags and not to the instantaneous time step. Furthermore, correlation may be interpreted as the square root of $R^{2}$ of a linear regression model, where $R^{2}$ represents the variance of the response variable which may be explained by the regressors, i.e., when partial CCA is applied between the past values of one subset and present values of another subset, it is identifying Granger causality between groups of time series.\pb

	In order to calculate $\rho$, set:
	
\begin{equation}
	{\bf u}={\bf a}'{\bf Y}_{t}^{i},
\end{equation}

and

\begin{equation}
	{\bf v}={\bf b}'{\bf Y}_{t-1}^{j},
\end{equation}

for some pair of coefficient vectors ${\bf a}$ and ${\bf b}$. Then, we obtain

\begin{equation}
	Var({\bf u})={\bf a}'Cov({\bf Y}_{t}^{i}){\bf a}={\bf a}'{\bf \Sigma}_{{\bf Y}_{t}^{i}{\bf Y}_{t}^{i}}{\bf a},
\end{equation}

\begin{equation}
	Var({\bf v})={\bf b}'Cov({\bf Y}_{t-1}^{j}){\bf b}={\bf b}'{\bf \Sigma}_{{\bf Y}_{t-1}^{j}{\bf Y}_{t-1}^{j}}{\bf b},
\end{equation}

\begin{equation}
	Cov({\bf u},{\bf v})={\bf a}'Cov({\bf Y}_{t}^{i},{\bf Y}_{t-1}^{j}){\bf b}={\bf a}'{\bf \Sigma}_{{\bf Y}_{t}^{i}{\bf Y}_{t-1}^{j}}{\bf b}.
\end{equation}

	We shall seek coefficient vectors ${\bf a}$ and ${\bf b}$ such that

\begin{equation}
	Corr({\bf u},{\bf v})=\frac{{\bf a}'{\bf \Sigma}_{{\bf Y}_{t}^{i}{\bf Y}_{t-1}^{j}}{\bf b}}{\sqrt{{\bf a}'{\bf \Sigma}_{{\bf Y}_{t}^{i}{\bf Y}_{t}^{i}}{\bf a}}\sqrt{{\bf b}'{\bf \Sigma}_{{\bf Y}_{t-1}^{j}{\bf Y}_{t-1}^{j}}{\bf b}}}
\end{equation}

is maximized.\pb

	However, notice that the calculations performed above do not identify conditional Granger causalities, i.e., Granger causalities partialized by a third set of time series. Therefore, it is necessary to develop a partial canonical correlation analysis.\pb
	
	The crucial point in performing partial canonical correlation analysis \cite{Rao69} is to derive from variance-covariance matrices, linear coefficients for combining original variables into canonical variables. It turns out that the residualized variance-covariance matrices of ${\bf Y}_{t}^{i}$ and ${\bf Y}_{t-1}^{j}$ after partialing out the effect of vector ${\bf X}={\bf Y}_{t-1}\backslash \{{\bf Y}_{t-1}^{j}\}$ from both ${\bf Y}_{t}^{i}$ and ${\bf Y}_{t-1}^{j}$, provide the solutions. More specifically, suppose that, when combined, the three vectors of variables, ${\bf Y}_{t}^{i}$, ${\bf Y}_{t-1}^{j}$ and ${\bf X}$ have the following partitioned variance-covariance matrix:

\begin{equation}
	\mathbf{{\bf \hat{\Sigma}}_{{\bf y}_{t}^{i}{\bf y}_{t-1}^{j}{\bf x}}}= \left( \begin{array}{ccc}
									{\bf \hat{\Sigma}}_{{\bf y}_{t}^{i}{\bf y}_{t}^{i}} & {\bf \hat{\Sigma}}_{{\bf y}_{t}^{i}{\bf y}_{t-1}^{j}} & {\bf \hat{\Sigma}}_{{\bf y}_{t}^{i}{\bf x}}\\
									{\bf \hat{\Sigma}}_{{\bf y}_{t-1}^{j}{\bf y}_{t}^{i}} & {\bf \hat{\Sigma}}_{{\bf y}_{t-1}^{j}{\bf y}_{t-1}^{j}} & {\bf \hat{\Sigma}}_{{\bf y}_{t-1}^{j}{\bf x}}\\
									{\bf \hat{\Sigma}}_{{\bf x}{\bf y}_{t}^{i}} & {\bf \hat{\Sigma}}_{{\bf x}{\bf y}_{t-1}^{j}} & {\bf \hat{\Sigma}}_{{\bf x}{\bf x}}\\
									\end{array} \right)
	\end{equation}
	
	where the $(r,s)$th entry of ${\bf \hat{\Sigma}}_{{\bf UV}}$ is given by $(T-1)^{-1}\sum_{k=1}^{T}({\bf U}_{kr}-{\bf \bar{U}}_{r})({\bf V}_{ks}-{\bf \bar{V}}_{s})$ ($T$: time series' length).\pb
	
	The conditional variance-covariance matrix of ${\bf y}_{t}^{i}$ and ${\bf y}_{t-1}^{j}$, partialing out the effect of ${\bf x}$ is given as \cite{Anderson84, Johnson02, Timm75}:

	\begin{equation}
	\mathbf{{\bf \hat{\Sigma}}_{{\bf y}_{t}^{i}{\bf y}_{t-1}^{j}|{\bf x}}}= \left( \begin{array}{cc}
									{\bf \hat{\Sigma}}_{{\bf y}_{t}^{i}{\bf y}_{t}^{i}|{\bf x}} & {\bf \hat{\Sigma}}_{{\bf y}_{t}^{i}{\bf y}_{t-1}^{j}|{\bf x}}\\
									{\bf \hat{\Sigma}}_{{\bf y}_{t-1}^{j}{\bf y}_{t}^{i}|{\bf x}} & {\bf \hat{\Sigma}}_{{\bf y}_{t-1}^{j}{\bf y}_{t-1}^{j}|{\bf x}}\\
									\end{array} \right),
	\end{equation}

	where
	\begin{eqnarray}
	{\bf \hat{\Sigma}}_{{\bf y}_{t}^{i}{\bf y}_{t}^{i}|{\bf x}} & = & {\bf \hat{\Sigma}}_{{\bf y}_{t}^{i}{\bf y}_{t}^{i}}-{\bf \hat{\Sigma}}_{{\bf y}_{t}^{i}{\bf x}}{\bf \hat{\Sigma}}_{{\bf x}{\bf x}}^{-1}{\bf \hat{\Sigma}}_{{\bf x}{\bf y}_{t}^{i}}\\
	{\bf \hat{\Sigma}}_{{\bf y}_{t}^{i}{\bf y}_{t-1}^{j}|{\bf x}} & = & {\bf \hat{\Sigma}}_{{\bf y}_{t}^{i}{\bf y}_{t-1}^{j}}-{\bf \hat{\Sigma}}_{{\bf y}_{t}^{i}{\bf x}}{\bf \hat{\Sigma}}_{{\bf x}{\bf x}}^{-1}{\bf \hat{\Sigma}}_{{\bf x}{\bf y}_{t-1}^{j}}\\
 	{\bf \hat{\Sigma}}_{{\bf y}_{t-1}^{j}{\bf y}_{t}^{i}|{\bf x}} & = & {\bf \hat{\Sigma}}_{{\bf y}_{t-1}^{j}{\bf y}_{t}^{i}}-{\bf \hat{\Sigma}}_{{\bf y}_{t-1}^{j}{\bf x}}{\bf \hat{\Sigma}}_{{\bf x}{\bf x}}^{-1}{\bf \hat{\Sigma}}_{{\bf x}{\bf y}_{t}^{i}}\\
 	{\bf \hat{\Sigma}}_{{\bf y}_{t-1}^{j}{\bf y}_{t-1}^{j}|{\bf x}} & = & {\bf \hat{\Sigma}}_{{\bf y}_{t-1}^{j}{\bf y}_{t-1}^{j}}-{\bf \hat{\Sigma}}_{{\bf y}_{t-1}^{j}{\bf x}}{\bf \hat{\Sigma}}_{{\bf x}{\bf x}}^{-1}{\bf \hat{\Sigma}}_{{\bf x}{\bf y}_{t-1}^{j}}
  \end{eqnarray}
  
	Similar to regular canonical correlation analysis \cite{Johnson02}, the eigenvalues $\lambda_{d}$ ($d=1,\ldots, \min(m,n)$) from the following two matrices ${\bf A}$ and ${\bf B}$, will be the squared partial canonical correlation coefficients for the $d$th canonical functions and the eigenvectors ${\bf a}_{d}$ and ${\bf b}_{d}$ associated with the eigenvalue $\lambda_{d}$ will be the linear coefficient vectors which combine the original variables into synthetic canonical variables:

\begin{equation}
	{\bf A}={\bf \hat{\Sigma}}_{{\bf y}_{t}^{i}{\bf y}_{t}^{i}|{\bf x}}^{-1/2}{\bf \hat{\Sigma}}_{{\bf y}_{t}^{i}{\bf y}_{t-1}^{j}|{\bf x}}{\bf \hat{\Sigma}}_{{\bf y}_{t-1}^{j}{\bf y}_{t-1}^{j}|{\bf x}}^{-1}{\bf \hat{\Sigma}}_{{\bf y}_{t-1}^{j}{\bf y}_{t}^{i}|{\bf x}}{\bf \hat{\Sigma}}_{{\bf y}_{t}^{i}{\bf y}_{t}^{i}|{\bf x}}^{-1/2}
\end{equation}

\begin{equation}
	{\bf B}={\bf \hat{\Sigma}}_{{\bf y}_{t-1}^{j}{\bf y}_{t-1}^{j}|{\bf x}}^{-1/2}{\bf \hat{\Sigma}}_{{\bf y}_{t-1}^{j}{\bf y}_{t}^{i}|{\bf x}}{\bf \hat{\Sigma}}_{{\bf y}_{t}^{i}{\bf y}_{t}^{i}|{\bf x}}^{-1}{\bf \hat{\Sigma}}_{{\bf y}_{t}^{i}{\bf y}_{t-1}^{j}|{\bf x}}{\bf \hat{\Sigma}}_{{\bf y}_{t-1}^{j}{\bf y}_{t-1}^{j}|{\bf x}}^{-1/2}
\end{equation}.

	Let $\lambda_{1} \geq \lambda_{2} \geq \ldots \geq \lambda_{\textrm{min($m$,$n$)}}$ be the ordered eigenvalues of matrices ${\bf A}$ and ${\bf B}$, therefore $CCA({\bf y}_{t}^{i},{\bf y}_{t-1}^{j}|{\bf y}_{t}\backslash \{{\bf y}_{t-1}^{j}\})=\hat{\rho}=\sqrt{\lambda_{1}}$.\pb
\indent Again, as in regular canonical correlation analysis, the two matrices ${\bf A}$ and ${\bf B}$, have the same eigenvalues but with different eigenvectors, i.e, each eigenvector ${\bf b}_{d}$ is proportional to ${\bf \hat{\Sigma}}_{{\bf y}_{t-1}^{j}{\bf y}_{t-1}^{j}|{\bf x}}^{-1/2}{\bf \hat{\Sigma}}_{{\bf y}_{t-1}^{j}{\bf y}_{t}^{i}|{\bf x}}{\bf \hat{\Sigma}}_{{\bf y}_{t}^{i}{\bf y}_{t}^{i}|{\bf x}}^{-1/2}{\bf a}_{d}$.

	\subsection*{Statistical test}
  The hypothesis to be tested in order to verify the existence of Granger causality between {\it sets} of variables is as follows:\pb

\begin{math}
	\indent H_{0}: CCA({\bf Y}_{t}^{i},{\bf Y}_{t-1}^{j}|{\bf X})=\rho=0 \indent \textrm{(Granger noncausality)}
\end{math}

\begin{math}
	H_{1}: CCA({\bf Y}_{t}^{i},{\bf Y}_{t-1}^{j}|{\bf X})=\rho \neq0 \indent \textrm{(Granger causality)}
\end{math}

	The bootstrap for our method is based on the block bootstrap \cite{Lahiri03}. It consists of splitting the data into blocks of observations (with overlapping) and sampling the blocks randomly with replacement. To describe this bootstrap more precisely, let the data consist of observations ${\bf Y}_{t}^{i}$ and ${\bf Y}_{t}^{j}$ ($t=1,\ldots,T$).\pb
	
	With overlapping blocks of length $l$, block 1 is observations ${{\bf Y}_{h}:h=1,\ldots,l}$, block 2 is observations ${{\bf Y}_{h+1}:h=1,\ldots,l}$, block 3 is observations ${{\bf Y}_{h+2}:h=1,\ldots,l}$ and so forth. The bootstrap sample ${\bf Y}_{t}^{i*}$ is obtained by sampling blocks randomly with replacement from ${\bf Y}_{t}^{i}$ and laying them end-to-end in the order sampled. The bootstrap sample ${\bf Y}_{t}^{j*}$ is obtained in an analogous way. This block re-sampling is carried out in order to capture the dependence structure of neighbourhood obsevations, i.e., autocorrelation. Moreover, the resampling of all time series of $Y_{t}^{i}$ is carried out together, in order to capture contemporaneous correlations between time series. The same is performed for the time series of $Y_{t}^{j}$. However, ${\bf Y}_{t}^{i}$ and ${\bf Y}_{t}^{j}$ are re-sampled independently, in order to break the relationship between the response and predictor variables.\pb
	
	After constructing the bootstrap samples ${\bf Y}_{t}^{i*}$ and ${\bf Y}_{t}^{j*}$, calculate $CCA({\bf Y}_{t}^{i*},{\bf Y}_{t-1}^{j*}|{\bf X}^{*})=\hat{\rho}^{*}$, where ${\bf X}^{*}={\bf Y}_{t-1}^{*}\backslash \{{\bf Y}_{t-1}^{j*}\}$. Repeat these steps until obtaining the desired number of bootstraps. Then, use the empirical distribution of $\rho^{*}$ to test whether $\hat{\rho}=0$ (the bootstrap scheme is illustrated in Figure 4).\pb
	
	Regardless of the block bootstrap that is used, the block length $l$ must increase with increasing time series length $T$ to render bootstrap estimators of moments and distribution functions consistent \cite{Carlstein86, Kunsch89, Hall85}. Similarly, the block length must increase with increasing sample size to enable the block bootstrap to achieve asymptotically correct coverage probabilities for confidence intervals and rejection probabilities for hypothesis tests. For the special case of an autoregressive process of order one, \cite{Carlstein86} showed that the block length $l$ that minimizes the asymptotic mean-square error of the variance estimator increases at the rate of $l\propto T^{\frac{1}{3}}$. Since time series gene expression data are generally short, it is unfeasible to fit an AR model of higher orders. However, if a longer time series data becomes available, one may use the algorithm proposed by \cite{Buhlmann99} in order to select the block length for the bootstrap procedure.

  \subsection*{Implementation}
	The method to identify Granger causality between sets of time series was implemented in R \cite{R}. The R code may be accessed in the Supplementary Material.
  
  \subsection*{Computational simulation}
	Before analyzing actual gene expression data, we conducted Monte Carlo simulations to examine whether our approach is able to identify Granger causality between sets of time series and effectively control the rate of false positives.\pb
	
	The artificial network is set as:
	
\begin{displaymath}
\mbox{simulation 1}\left\{\begin{array}{l}
	Z_{1,t}^{\textrm{{\scriptsize I}}}    = \beta \times Z_{1,t-1}^{\textrm{{\scriptsize I}}} - \beta \times Z_{4,t-1}^{\textrm{{\scriptsize I}}} + \varepsilon_{1,t}\\
	Z_{2,t}^{\textrm{{\scriptsize I}}}    = \beta \times Z_{1,t-1}^{\textrm{{\scriptsize I}}} + \varepsilon_{2,t}^{\textrm{{\scriptsize I}}}\\
	Z_{3,t}^{\textrm{{\scriptsize I}}}    = \beta \times Z_{1,t-1}^{\textrm{{\scriptsize I}}} - \beta \times Z_{5,t-1}^{\textrm{{\scriptsize I}}} + \varepsilon_{3,t}\\
	Z_{4,t}^{\textrm{{\scriptsize I}}}    = \beta \times Z_{2,t-1}^{\textrm{{\scriptsize I}}} + \varepsilon_{4,t}^{\textrm{{\scriptsize I}}}\\
	Z_{5,t}^{\textrm{{\scriptsize I}}}    = \varepsilon_{5,t}\\
	Z_{6,t}^{\textrm{{\scriptsize II}}}   = \beta \times Z_{8,t-1}^{\textrm{{\scriptsize II}}} + \varepsilon_{6,t}^{\textrm{{\scriptsize II}}}\\
	Z_{7,t}^{\textrm{{\scriptsize II}}}   = \beta \times Z_{3,t-1}^{\textrm{{\scriptsize I}}} - \beta \times Z_{6,t-1}^{\textrm{{\scriptsize II}}} + \varepsilon_{7,t}\\
	Z_{8,t}^{\textrm{{\scriptsize II}}}   = \beta \times Z_{10,t-1}^{\textrm{{\scriptsize II}}} + \varepsilon_{8,t}\\
	Z_{9,t}^{\textrm{{\scriptsize II}}}   =  \beta \times Z_{5,t-1}^{\textrm{{\scriptsize I}}} - \beta \times Z_{7,t-1}^{\textrm{{\scriptsize II}}} + \beta \times Z_{14,t-1}^{\textrm{{\scriptsize III}}} + \varepsilon_{9,t}\\
	Z_{10,t}^{\textrm{{\scriptsize II}}}  =  \beta \times Z_{9,t-1}^{\textrm{{\scriptsize II}}} - \beta \times Z_{13,t-1}^{\textrm{{\scriptsize III}}} + \varepsilon_{10,t}\\
	Z_{11,t}^{\textrm{{\scriptsize III}}} = \beta \times Z_{8,t-1}^{\textrm{{\scriptsize II}}} + \varepsilon_{11,t}\\
	Z_{12,t}^{\textrm{{\scriptsize III}}} = \varepsilon_{12,t}\\
 	Z_{13,t}^{\textrm{{\scriptsize III}}} = \varepsilon_{13,t}\\
	Z_{14,t}^{\textrm{{\scriptsize III}}} = \varepsilon_{14,t}\\
	\end{array}\right.
\end{displaymath}

for $t=1,\ldots,T$, where the noises $\varepsilon_{i,t}$ $(i=1,\ldots,14)$ are normally distributed with mean of zero and variance of one and I, II and III represent the sets to which each time series $Z_{i,t}$ belong. The time series length is set to 100, i.e., $T=100$, and $\beta$ is set to 0.4.\\
\indent In order to evaluate the power in identifying Granger causalities between sets of time series, a comparison between the proposed method based on CCA and the standard VAR model was performed. The following artificial network is used for the comparison:

\begin{displaymath}
\mbox{simulation 2}\left\{\begin{array}{l}
	W_{1,t}^{\textrm{{\scriptsize I}}}    = \gamma \times W_{1,t-1}^{\textrm{{\scriptsize I}}} + \epsilon_{1,t}\\
	W_{2,t}^{\textrm{{\scriptsize I}}}    = \gamma \times W_{1,t-1}^{\textrm{{\scriptsize I}}} - \gamma \times W_{5,t-1}^{\textrm{{\scriptsize I}}} + \epsilon_{2,t}\\
	W_{3,t}^{\textrm{{\scriptsize I}}}    = \gamma \times W_{1,t-1}^{\textrm{{\scriptsize I}}} - \gamma \times W_{4,t-1}^{\textrm{{\scriptsize I}}} + \epsilon_{3,t}\\
	W_{4,t}^{\textrm{{\scriptsize I}}}    = \gamma \times W_{2,t-1}^{\textrm{{\scriptsize I}}} - \gamma \times W_{4,t-1}^{\textrm{{\scriptsize I}}} + \epsilon_{4,t}\\
	W_{5,t}^{\textrm{{\scriptsize I}}}    = \gamma \times W_{3,t-1}^{\textrm{{\scriptsize I}}} - \gamma \times W_{13,t-1}^{\textrm{{\scriptsize III}}} + \epsilon_{5,t}\\
	W_{6,t}^{\textrm{{\scriptsize II}}}   = \gamma \times W_{3,t-1}^{\textrm{{\scriptsize I}}} + \epsilon_{6,t}\\
	W_{7,t}^{\textrm{{\scriptsize II}}}   = \gamma \times W_{3,t-1}^{\textrm{{\scriptsize I}}} - \gamma \times W_{6,t-1}^{\textrm{{\scriptsize II}}} + \epsilon_{7,t}\\
	W_{8,t}^{\textrm{{\scriptsize II}}}   = \gamma \times W_{3,t-1}^{\textrm{{\scriptsize I}}} - \gamma \times W_{7,t-1}^{\textrm{{\scriptsize II}}} + \epsilon_{8,t}\\
	W_{9,t}^{\textrm{{\scriptsize II}}}   = \gamma \times W_{3,t-1}^{\textrm{{\scriptsize I}}} - \gamma \times W_{8,t-1}^{\textrm{{\scriptsize II}}} + \epsilon_{9,t}\\
	W_{10,t}^{\textrm{{\scriptsize II}}}  = \gamma \times W_{3,t-1}^{\textrm{{\scriptsize I}}} - \gamma \times W_{9,t-1}^{\textrm{{\scriptsize II}}} + \gamma \times W_{12,t-1}^{\textrm{{\scriptsize III}}} + \epsilon_{10,t}\\
	W_{11,t}^{\textrm{{\scriptsize III}}}  = \gamma \times W_{11,t-1}^{\textrm{{\scriptsize III}}} - \gamma \times W_{13,t-1}^{\textrm{{\scriptsize III}}} + \epsilon_{11,t}\\
	W_{12,t}^{\textrm{{\scriptsize III}}}  = \gamma \times W_{11,t-1}^{\textrm{{\scriptsize III}}} + \epsilon_{12,t}\\
	W_{13,t}^{\textrm{{\scriptsize III}}}  = \gamma \times W_{12,t-1}^{\textrm{{\scriptsize III}}} + \epsilon_{13,t}\\
	\end{array}\right.
\end{displaymath}

	for $t=1,\ldots,T$, where the noises $\epsilon_{i,t}$ $(i=1,\ldots,13)$ are normally distributed with mean of zero and variance of one and I, II and III represent the sets which each time series $W_{i,t}$ belongs to. The time series length is set to 100, i.e., $T=100$, and $\gamma$ is set to 0.2.\pb

	For each simulation, 1 and 2, 10,000 networks were generated. For each artificial network, the proposed approach was applied to identify Granger causalities between the sets of variables. The threshold to discriminate the existence of an edge was $p < 0.05$, i.e., the false positives rate was controlled to $5\%$ and the number of bootstraps was set to 1,000. For the standard VAR model, the Wald's test was used in order to identify and test Granger causality between sets. In other words, suppose three sets I, II and III as in simulation 2. The coefficients of the VAR model related to the time series of set I to II were simultaneously tested in order to verify whether they are equal to zero or not. The same was performed with the coefficients I $\rightarrow$ I, I $\rightarrow$ III, II $\rightarrow$ I, II $\rightarrow$ II, III $\rightarrow$ I, III $\rightarrow$ II and III $\rightarrow$ III.

\section*{Authors contributions}
    AF has made substantial contributions to the conception, design and implementation of the study, and has also been responsible for drafting the manuscript. JRS and KK have made substantial contributions to data analysis and application of statistical concepts. LRG has made substantial contributions to the biological interpretations, and has been responsible for drafting some parts of the manuscript. MN, MCS and SM have discussed the results and critically revised the manuscript for important intellectual content.

\section*{Acknowledgements}
  This work was supported by grants of RIKEN and by Brazilian funding agencies (FAPESP, CNPq, FINEP, CAPES).


{\ifthenelse{\boolean{publ}}{\footnotesize}{\small}
 \bibliographystyle{bmc_article}  
  \bibliography{bmc_article} }     


\ifthenelse{\boolean{publ}}{\end{multicols}}{}



  \subsection*{Figure 1 - Simulation 1}
      Simulation 1. (A) Time series data for each simulated variable; (B) Time series $Z_{i}$ ($i=1,\ldots,14$) represent variables. I, II and III are the sets of time series composed by $Z_{i}$. Arrows represent the Granger causalities; (C) Granger causality between sets of time series.

  \subsection*{Figure 2 - Simulation 2}
      Simulation 2. (A) The time series $W_{i}$ ($i=1,\ldots,13$) represent the variables. I, II and III are the sets of time series composed by $W_{i}$. Arrows represent the Granger causalities; (B) Granger causality between sets of time series.

  \subsection*{Figure 3 - Actual biological network}
      Identification of Granger causality to a network composed by I: RECK, SRC, C-Myc, TIMP2; II: TP53, p21, GADD45A; III: FGFRL1, FGF2, FGFR2, FGFR1 and IV: NEMO, IKKA, IKB and NFKB genes. Connectivities inside each set were identified by using the standard VAR model. Dashed arrows are significant connectivities with $p<0.1$ and full arrows are significant connectivities with $p<0.05$.

  \subsection*{Figure 4 - Illustrative diagram of the bootstrap.}
      Time series 1, 2 and 3 belongs to the set $Y_{t}^{i}$ while time series 4 and 5 belong to the $Y_{t}^{j}$ set. Notice that the blocks may be overlapped.


\section*{Tables}
  
  \subsection*{Table 1 - Simulation 1. Number of Granger causalities obtained using our approach in 10,000 simulations. The rate of false positives was controlled in 5\%, i.e., it is expected to obtain $\sim500$ false-positives where there is no Granger causality. In bold, are the Granger causalities which actually exist.} \label{Tab:01}
    \par \mbox{}
    \par
    \mbox{
      \begin{tabular}{cccccccccccccccc}
      \hline
			from/to &&&&& I &&&&& II &&&&& III\\
			\hline
			I &&&&& \bf{10,000} &&&&& \bf{8,543} &&&&& 507\\
			II &&&&& 461 &&&&& \bf{9,979} &&&&& \bf{6,646}\\
			III &&&&& 500 &&&&& \bf{8,015} &&&&& 437\\
			\hline
      \end{tabular}
      }

\subsection*{Table 2 - Simulation 2. Number of Granger causalities obtained using our approach in 10,000 simulations. In parenthesis are the number of Granger causalities obtained by standard VAR. The rate of false positives was controlled in 5\%, i.e., it is expected to obtain $\sim500$ false-positives where there is no Granger causality. In bold, are the Granger causalities which actually exist.} \label{Tab:01}
    \par \mbox{}
    \par
    \mbox{
      \begin{tabular}{cccccccccccccccc}
      \hline
			from/to &&&&& I &&&&& II &&&&& III\\
			\hline
			I &&&&& \bf{7,154} ({\bf 1,079}) &&&&& \bf{10,000} ({\bf 9,980}) &&&&& 505 (589)\\
		  II &&&&& 466 (547) &&&&& \bf{10,000} ({\bf 9,997}) &&&&& 463 (522)\\
		  III &&&&& \bf{1,490} ({\bf 1,015}) &&&&& {\bf 10,000} ({\bf 9,827}) &&&&& {\bf 6,263} ({\bf 2,107})\\
			\hline
      \end{tabular}
      }


\section*{Additional Files}
  \subsection*{Additional file 1 --- R code}
    R code to identify Granger causality between gene sets.

\end{bmcformat}
\end{document}